\documentclass{birkjour_t2}
\usepackage{color}

 \theoremstyle{definition}
 
 \theoremstyle{remark}
 
\DeclareMathOperator{\sech}{sech}
\usepackage{subfigure}

\begin{document}

\title{Deformed two-dimensional rogue waves in the $(2+1)$-dimensional Korteweg-de Vries equation}

\author{Yulei Cao$^1$}

\address{$^1$ Institute for Advanced Study, Shenzhen University, Shenzhen,
Guangdong 518060, P.\ R.\ China \\}

\email{caoyulei@mail.ustc.edu.cn}

\thanks{$^*$ Corresponding author: pyhu@szu.edu.cn}
\author{Peng-Yan Hu$^{2*}$}
\address{$^{2*}$ College of Mathematics and Statistics, Shenzhen University, Shenzhen 518060, China \\}
\email{pyhu@szu.edu.cn}

\author{Yi Cheng$^{3}$}
\address{$^3$ School of Mathematical Sciences, USTC, Hefei, Anhui 230026, P.\ R.\ China \\}
\email{chengy@ustc.edu.cn}

\author{Jingsong He$^{1}$}

\address{$^{1}$ Institute for Advanced Study, Shenzhen University, Shenzhen,
Guangdong 518060, P.\ R.\ China \\}
\email{hejingsong@nbu.edu.cn; jshe@ustc.edu.cn}

\begin{abstract}
Within the $(2+1)$-dimensional Korteweg-de Vries equation framework, new
bilinear B$\rm\ddot{a}$cklund transformation and Lax pair
are presented based on the binary Bell polynomials and gauge transformation.
By introducing an arbitrary function $\phi(y)$, a family of
deformed soliton and deformed breather solutions are presented with the
improved Hirota's bilinear method. Choosing the appropriate
parameters, their interesting dynamic behaviors are
shown in three-dimensional plots. Furthermore, novel rational solutions
are generated by taking the limit of obtained solitons. Additionally,
two dimensional [2D] rogue waves (localized in both space and time) on the soliton plane are presented,
we refer to it as deformed 2D rogue waves.
The obtained deformed 2D rogue waves can be viewed as a 2D
analog of the Peregrine soliton on soliton plane, and its evolution process is
analyzed in detail. The deformed 2D rogue wave solutions are
constructed successfully, which are closely related to the
arbitrary function $\phi(y)$. This new idea is also applicable to
other nonlinear systems.
\end{abstract}

\maketitle


\vspace{-0.9cm}




\section{Introduction}
Rogue waves (RWs) in the ocean, also known as killer waves,
monster waves, freak waves and extreme waves, are special waves
with extremely destructive, which should be responsible for some
maritime disasters. The main feature of RW is that "appear from nowhere and
disappear without a trace"\cite{RW}. Peregrine was the first one who obtained
the RW solution from one-dimensional systems\cite{dh}, so such RW solution is also called
"Peregrine soliton". These RWs are localized in both space and time
in one-dimensional systems\cite{guo}-\cite{he5}, and its dynamic behaviors are similar to
that of lumps in high-dimensional systems\cite{jg1}-\cite{jg4}. RWs in high-dimensional
systems are often called line RWs, which are merely localized in time\cite{jg5}-\cite{jg8}.
The study of RW has been widely used in diverse areas of theoretical
and applied physics, including plasma physics\cite{MMW1,MMW2},
Bose-Einstein condensates\cite{BKA1,BKA2}, atmosphere physics\cite{an2},
optics and photonics\cite{MBR1,MBR2,MBR3} and superfluids\cite{an1}.

Now, a fundamental problem occurs: Is it possible that other types of RW solutions
exist in nonlinear systems? As well as seeking new exact solution for
nonlinear systems is an open problem and a challenging work. Furthermore,
in integrable system, the Peregrine soliton only appears in the
one-dimensional systems. But, the RWs, which are localized
in time and space, have never been found in the high-dimensional system.
Therefore, searching for the RW solution that is local in time and space of high-dimensional
systems has always been an open problem. Until recently, Guo et al.\cite{glj1} obtained
RW solutions, which are localized in both space and time, from a two-dimensional nonlinear
Schr$\rm\ddot{o}$dinger model by the even-fold Darboux transformation.
Now the long-standing problem of the construction of two-dimensional [2D]
rogue wave on zero background is addressed\cite{glj1}.
This is a pioneering work, which inspires us to explore the structure
of 2D RW solutions of other high-dimensional soliton equations.
Additionally, RW solutions of complex nonlinear equations have been widely
researched, but the RW solutions of real nonlinear equations are rarely mentioned,
which is another motivation.

Inspired by the above considerations, in this paper, we focus on the
$(2+1)$-dimensional [2D] Korteweg-de Vries [KdV] equation
\begin{equation} \label{N1}
\begin{aligned}
U_{t}+U_{xxx}-3(UV)_{x}=0,\qquad U_{x}=V_{y},
\end{aligned}
\end{equation}
which was first derived by Boiti-Leon-Manna-Pempinelli with a
weak Lax pair\cite{BLMP}, and can be reduced to the celebrated
KdV equation if $x=y$. Equation\eqref{N1} is an asymmetric part
of the Nizhnik-Novikov-Veselov [NNV] equation\cite{Lou}, thus equation\eqref{N1}
is also called ANNV equation; this equation\eqref{N1} can also be derived
by using the inner parameter-dependent symmetry constraint of the KP equation\cite{Lou},
and can be considered as a model of an incompressible
fluid where $U$ and $V$ are the component of velocity\cite{est}.
Equation\eqref{N1} was also regarded to be a generalization\cite{lu1}
of the results of Hirota and Satsuma\cite{lu2} in the $(2 + 1)$ dimension.

Letting $U=W_y, V=W_{x}$, then equation\eqref{N1} reduce to
the following single equation,
\begin{equation}\label{N2}
\begin{aligned}
W_{yt}+W_{xxxy}-3W_{xx}W_{y}-3W_{x}W_{xy}=0.
\end{aligned}
\end{equation}
This equation\eqref{N2} known as BLMP equation was widely
investigated. The Cauchy problem, associated with initial
data decaying sufficiently rapidly at infinity, was solved
by means of inverse scattering transformation\cite{BLMP}.
The spectral transformation, nonclassical symmetries and
Painlev$\acute{e}$ property for equation\eqref{N2} has been
researched in Refs.\cite{BLMP,ab,pa}.
And a series of B$\rm\ddot{a}$cklund transformations\cite{back1,back2,back3},
Lax pairs\cite{back3},
supersymmetric\cite{super}, soliton-like\cite{t1}-\cite{chen},
breather\cite{fan,luo} and lump-type\cite{csl,wxy,wcj} solutions
are derived.

In this letter, we are committed to exploring new B$\rm\ddot{a}$cklund
transformations, Lax pairs and new RW solutions of 2D KdV
equation\eqref{N1} by the Hirota method.
The new RW solution is called deformed 2D RW because its formula involves an
arbitrary function $\phi(y)$ which provides a deformed background for rogue wave.
This arbitrary function $\phi(y)$ comes from the crucial bilinear transformation which maps
$U$ to a bilinear form.  In section\ref{1}, new B$\rm\ddot{a}$cklund transformations
and Lax pair are derived based on the binary Bell polynomials and linearizing
the bilinear equation. In section\ref{2},
we obtain deformed kink soliton and deformed breather
solutions of 2D KdV equation by means of the Hirota
method. In section\ref{3}, a family of new rational solutions and deformed 2D RW solutions
of the equation\eqref{N1} are presented by introducing an
arbitrary function $\phi(y)$.
The main results of the paper are
summarized and discussed in section\ref{4}.

\section{Bilinear B$\rm\ddot{a}$cklund transformation and associated Lax pair}\label{1}
In this section, we mainly focus on the new B$\rm\ddot{a}$cklund transformation
and Lax pair of 2D KdV equation\eqref{N1}. B$\rm\ddot{a}$cklund transformation and
lax pair are recognized as the main characteristics of integrability, which can be used to
obtain exact solutions of nonlinear systems. We first introduce
the following variables transformation
\begin{equation} \label{1s1}
\begin{aligned}
U=-q_{xy},\quad V=-q_{xx},
\end{aligned}
\end{equation}
then 2D KdV equation\eqref{N1} becomes
\begin{equation}
\begin{aligned}
E(q)=q_{yt}+q_{xxxy}+3q_{xx}q_{xy}=0,
\end{aligned}
\end{equation}
two different solutions $q=2\ln G$ and $\widetilde{q}=2\ln F$ of equation\eqref{N1} are introduced,
the corresponding two-field condition is as follows\cite{bt1}-\cite{bt5}
\begin{equation}\label{bk1}
\begin{aligned}
E(\widetilde{q})-E(q)=(\widetilde{q}-q)_{yt}+(\widetilde{q}-q)_{xxxy}
+3(\widetilde{q}_{xx}\widetilde{q}_{xy}-q_{xx}q_{xy})=0.
\end{aligned}
\end{equation}
We introduce two new variables
\begin{equation}
\begin{aligned}
v=\frac{\widetilde{q}-q}{2}=\ln\frac{F}{G},\quad w=\frac{\widetilde{q}+q}{2}=\ln FG.
\end{aligned}
\end{equation}
Based on the binary Bell polynomials, equation\eqref{bk1} can be rewritten into $\mathcal{Y}$-polynomials from Bell
polynomial theory as follows
and rewrite the equation\eqref{bk1} into the form (see Refs. \cite{bt3,bt4,bt5} for details)
\begin{equation}\label{bk2}
\begin{aligned}
E(\widetilde{q})-E(q)&=2[v_{yt}+v_{xxxy}+3w_{xx}v_{xy}+3w_{xy}v_{xx}]\\
&=2[\partial_y(\mathcal{Y}_{xxx}+\mathcal{Y}_{t})+3w_{xy}v_{xx}-3w_{xxy}v_{x}-3v_{x}^2v_{xy}].
\end{aligned}
\end{equation}
Through the following constraints
\begin{equation}
\begin{aligned}
w_{xy}+v_xv_y-\lambda v_x=0,
\end{aligned}
\end{equation}
where $\lambda$ is the constant parameter. Then, the following coupled system of $\mathcal{Y}$-polynomials of equation\eqref{N1} is deduced
\begin{equation}\label{bk3}
\begin{aligned}
\mathcal{Y}_{xy}-\lambda\mathcal{Y}_{x}=0,\\
\mathcal{Y}_{xxx}+\mathcal{Y}_{t}=0.
\end{aligned}
\end{equation}
According to the relationship between $\mathcal{Y}$-polynomials\cite{bt3,bt4,bt5} and bilinear operators $D$\cite{hirota}
\begin{equation}
\begin{aligned}
\mathcal{Y}_{n_1x_1,\cdots,n_lx_l}(v=\ln\frac{F}{G}, w=\ln FG) &=& (FG)^{-1}D_{x_1}^{n_1}\cdots
D_{x_l}^{n_l}F\cdot G,
\end{aligned}
\end{equation}
the following bilinear B$\rm\ddot{a}$cklund transformation is derived for the 2D KdV equation\eqref{N1}
\begin{equation}\label{bk4}
\begin{aligned}
(D_xD_y-\lambda D_x)F\cdot G=0,\\
(D^3_x+D_t)F\cdot G=0.
\end{aligned}
\end{equation}
Then logarithmic linearization of $\mathcal{Y}$-polynomials under the Hopf-Cole transformation $v=\ln\psi$,
and Bell polynomial formula
\begin{equation}
\begin{aligned}
&(FG)^{-1}D_{x_1}^{n_1}\cdots D_{x_l}^{n_l}F\cdot G\bigg{|}_{G=e^{\frac{q}{2}}, \frac{F}{G}=\psi}\\
=&\sum\limits_{r_1+\cdots+r_l=even}\sum\limits_{r_1=0}^{n_1}\cdots\sum\limits_{r_l=0}^{n_l}
\prod\limits_{i=1}^{l} \begin{pmatrix} n_i
\\ r_i \end{pmatrix}P_{r_1x_1\cdots r_lx_l}(q)\\
&\times \mathcal{Y}_{(n_1-r_1)x_1,\cdots,(n_l-r_l)x_l}(v),
\end{aligned}
\end{equation}
the $\mathcal{Y}$-polynomials can be written as
\begin{equation}\label{hopf}
\begin{aligned}
\mathcal{Y}_{n_1x_1,\cdots,n_lx_l}(v)=\frac{\psi_{n_1x_1,\cdots,n_lx_l}}{\psi},
\end{aligned}
\end{equation}
thus the system\eqref{bk3} is then linearized into a Lax pair
\begin{equation}
\begin{aligned}
&\psi_{xy}+q_{xy}\psi-\lambda\psi_{x}=0,\\
&\psi_{xxx}+3q_{xx}\psi_{x}+\psi_{t}=0,
\end{aligned}
\end{equation}
which is equivalent to the Lax pair of equation\eqref{N1}\cite{back1}
\begin{equation}
\begin{aligned}
&\psi_{xy}-U\psi-\lambda\psi_{x}=0,\\
&\psi_{xxx}-3V\psi_{x}+\psi_{t}=0.
\end{aligned}
\end{equation}
In order to get a new B$\rm\ddot{a}$cklund transformation, we introduce the following gauge transformation to the bilinear B$\rm\ddot{a}$cklund transformation\eqref{bk4}
\begin{equation}\label{gauge}
\begin{aligned}
F\rightarrow e^{\xi}F,\quad G\rightarrow e^{\eta}G,
\end{aligned}
\end{equation}
where $\xi=m_1x+n_1y+l_1t$ and $\eta=m_2x+n_2y+l_2t$. System\eqref{bk4} is transformed into
\begin{equation}
\begin{aligned}
\left[(D_x+m_1-m_2)(D_y+n_1-n_2)-\lambda (D_x+m_1-m_2)\right]F\cdot G=0,\\
\left[(D_x+m_1-m_2)^3+(D_t+l_1-l_2)\right]F\cdot G=0,
\end{aligned}
\end{equation}
under the following constraints
\begin{equation}
\begin{aligned}
m_1-m_2=\lambda, \quad n_1-n_2=\lambda,\quad l_2-l_1=\lambda^3.
\end{aligned}
\end{equation}
Then, a new B$\rm\ddot{a}$cklund transformation of equation\eqref{N1} is derived as follows
\begin{equation}\label{nbk1}
\begin{aligned}
(D_xD_y+\lambda D_y)F\cdot G=0,\\
(D^3_x+3\lambda D^2_x+3\lambda^2 D_x+D_t)F\cdot G=0.
\end{aligned}
\end{equation}
Similarly, the Lax pair of system\eqref{nbk1} is as follows
\begin{equation}
\begin{aligned}
\psi_{xy}+\lambda\psi_{y}+q_{xy}\psi=0,\\
\psi_{xxx}+2\lambda\psi_{xx}+3(\lambda^2+q_{xx})\psi_{x}+3\lambda q_{xx}\psi+\psi_t=0,
\end{aligned}
\end{equation}
or equivalently
\begin{equation}
\begin{aligned}
(L_1-U)\psi=&\psi_{xy}+\lambda\psi_{y}-U\psi=0,\\
(\partial_t+L_2)\psi=&\psi_{xxx}+2\lambda\psi_{xx}+3(\lambda^2-V)\psi_{x}-3\lambda V\psi+\psi_t=0.
\end{aligned}
\end{equation}
It is easy to check that the integrability condition $[L_1-U, \partial_t+L_2]\psi=0$ is satisfied.

Additionally, another form of B$\rm\ddot{a}$cklund transformation and lax pair of 2D KdV equation\eqref{N1} are also derived, when taking
\begin{equation} \label{s1}
\begin{aligned}
U=-2\partial_{xy} \ln f(x,y,t)+\phi(y),\quad V=-2\partial_{xx} \ln f(x,y,t),
\end{aligned}
\end{equation}
2D KdV equation\eqref{N1} is then translated into the following bilinear form,
\begin{equation}
\begin{aligned}\label{newsolu}
[D^{3}_{x}D_{y}+D_{y}D_{t}-3\phi(y)D^{2}_{x}]f(x,y,t) \cdot f(x,y,t) =0,
\end{aligned}
\end{equation}
$\phi(y)$ is
an arbitrary function of $y$, $f$ is a real function
and  $D$ is the Hirota's bilinear differential operator\cite{hirota},
based on the binary Bell polynomials\cite{bt1,bt2,bt3}, equation\eqref{N1} admits the following B$\rm\ddot{a}$cklund transformation
\begin{equation}\label{nnbk1}
\begin{aligned}
\left[D_xD_y-\phi(y)\right]F\cdot G=0,\\
\left[D^3_x+D_t+k\right]F\cdot G=0.
\end{aligned}
\end{equation}
If using gauge transformation\eqref{gauge} to the above bilinear B$\rm\ddot{a}$cklund transformation\eqref{nnbk1},
a new B$\rm\ddot{a}$cklund transformation of equation\eqref{N1} is derived
\begin{equation}\label{nbk1}
\begin{aligned}
\left[D_xD_y+\lambda(D_x+D_y)+\lambda^2-\phi(y)\right]F\cdot G=0,\\
\left[D^3_x+3\lambda D^2_x+3\lambda^2 D_x+D_t+k\right]F\cdot G=0.
\end{aligned}
\end{equation}
Then, we derive the corresponding linear system under the Hopf-Cole transformation\eqref{hopf}
\begin{equation}
\begin{aligned}
\psi_{xy}+\lambda(\psi_x+\psi_y)+(\lambda^2+q_{xy}-\phi(y))\psi=0,\\
\psi_{xxx}+3\lambda\psi_{xx}+3(\lambda^2+q_{xx})\psi_{x}+\psi_t+(3\lambda q_{xx}+k)\psi=0,
\end{aligned}
\end{equation}
which is equivalent to the Lax system of 2D KdV equation\eqref{N1}
\begin{equation}
\begin{aligned}
\psi_{xy}+\lambda(\psi_x+\psi_y)+(\lambda^2-U)\psi=0,\\
\psi_{xxx}+3\lambda\psi_{xx}+3(\lambda^2-V)\psi_{x}+\psi_t+(k-3\lambda V)\psi=0.
\end{aligned}
\end{equation}

\section{Deformed multi-kink soliton and deformed breather solutions of the 2D KdV equation}\label{2}
\begin{figure}[!htbp]
\centering
\subfigure[$\Phi(y)=\sech(y)$]{\includegraphics[height=3cm,width=3.5cm]{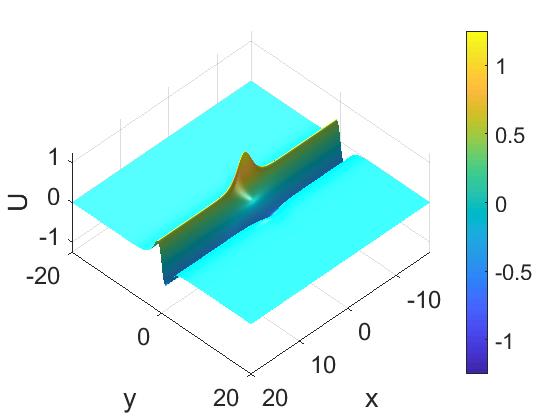}}
\subfigure[$\Phi(y)=\sech(y)$]{\includegraphics[height=3cm,width=3.5cm]{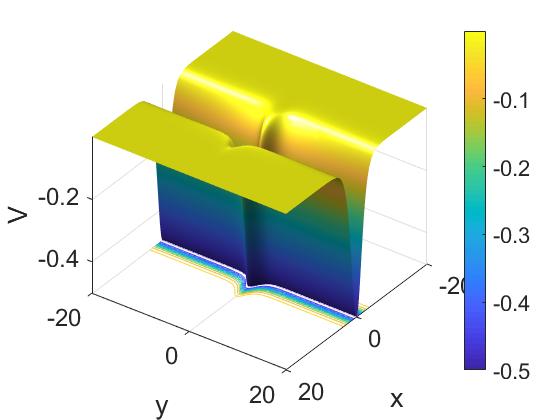}}
\subfigure[$\Phi(y)=\sin(y)$]{\includegraphics[height=3cm,width=3.5cm]{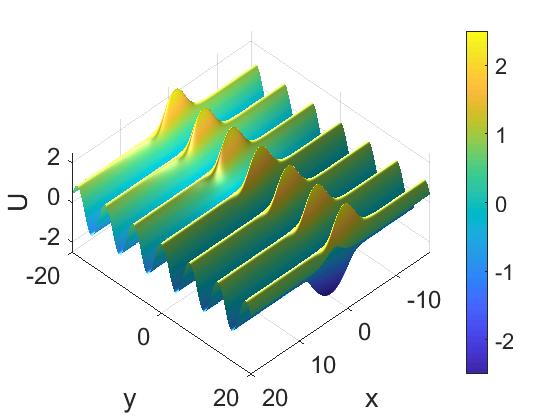}}
\subfigure[$\Phi(y)=\sin(y)$]{\includegraphics[height=3cm,width=3.5cm]{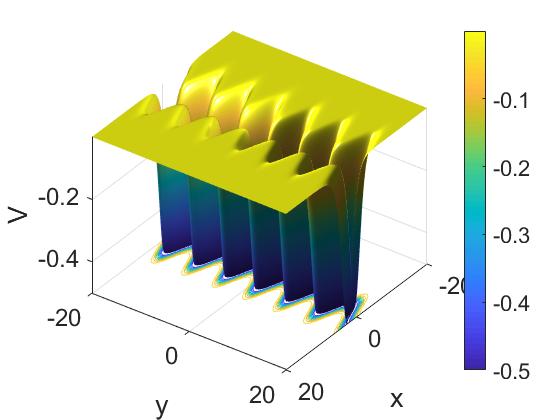}}
\caption{Deformed one-soliton $U_{[1]}$ and $V_{[1]}$ of equation\eqref{N1} in the $(x,y)$-plane with parameters
$p_1=1, q_1=-4$ and displayed at $t=0$.}\label{fig1}
\end{figure}

Based on the above new bilinear equation\eqref{newsolu}, some exact
solutions of 2D KdV equation\eqref{N1}, including deformed multi-solitons
and deformed breathers, are generated. First, deformed one-soliton solutions take the forms
\begin{equation}\label{u1}
\begin{aligned}
U_{[1]}=-2\partial_{xy}\ln(1+e^{\eta_1})+\phi(y),\quad V_{[1]}=-2\partial_{xx} \ln(1+e^{\eta_1}),
\end{aligned}
\end{equation}
where $\eta_1=p_{1}x+\Phi(y)q_{1}y+\Omega_{1}t+\eta^{0}_{1}, \;\eta^{0}_{1}=0$ and $\phi(y)=\frac{d\Phi(y)}{dy}$.
Substituting \eqref{u1} into \eqref{N1}, the following dispersion relation of 2D KdV equation is obtained
$$\Omega_{1}=-\frac{p_1^2(p_1q_1-3)}{q_1}.$$
The exact expression of $U_{[1]}$ and $V_{[1]}$ are as follows
\begin{equation}\label{1s}
\begin{aligned}
U_{[1]}&=\phi(y)\frac{1+\cosh\left(p_1x+q_1\Phi(y)+(\frac{3p^2_1}{q_1}-p^3_1)t
\right)-p_1q_1}{1+\cosh\left(p_1x+q_1\Phi(y)+(\frac{3p^2_1}{q_1}-p^3_1)t\right)},\\
V_{[1]}&=-\frac{p^2_1}{1+\cosh\left(p_1x+q_1\Phi(y)+(\frac{3p^2_1}{q_1}-p^3_1)t\right)}.
\end{aligned}
\end{equation}
As can be seen from the above expression, the extremum lines $p_1x+q_1\Phi(y)+(\frac{3p^2_1}{q_1}-p^3_1)t$ and velocity $\frac{3p_1-q_1p_1^2}{q_1}$ of $U_{[1]}$ and $V_{[1]}$ are the same, and the corresponding amplitudes of $U_{[1]}$ and $V_{[1]}$ in $(x,y)$-plane are $\phi(y)(1-\frac{p_1q_1}{2})$ and $\frac{p_1^2}{2}$ respectively.
Thus, $U_{[1]}$ is a deformed bright soliton when $\phi(y)(2-p_1q_1)>0$ and a deformed dark soliton when $\phi(y)(2-p_1q_1)<0$. However,
$V_{[1]}$ is always a deformed dark soliton see Fig.\ref{fig1}.

\begin{figure}[!htbp]
\centering
\subfigure[$\Phi(y)=\sech(y)$]{\includegraphics[height=2.5cm,width=3.5cm]{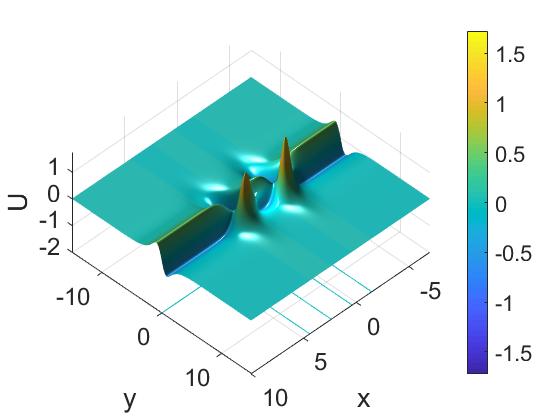}}
\subfigure[$\Phi(y)=\sech(y)$]{\includegraphics[height=2.5cm,width=3.5cm]{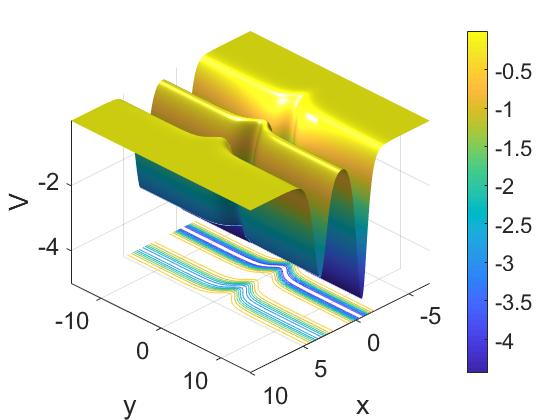}}
\subfigure[$\Phi(y)=\sin(y)$]{\includegraphics[height=2.5cm,width=3.5cm]{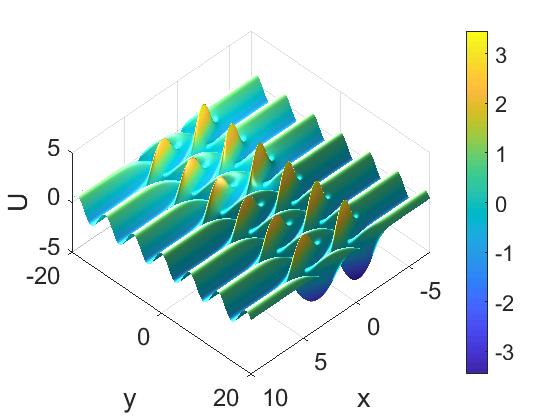}}
\subfigure[$\Phi(y)=\sin(y)$]{\includegraphics[height=2.5cm,width=3.5cm]{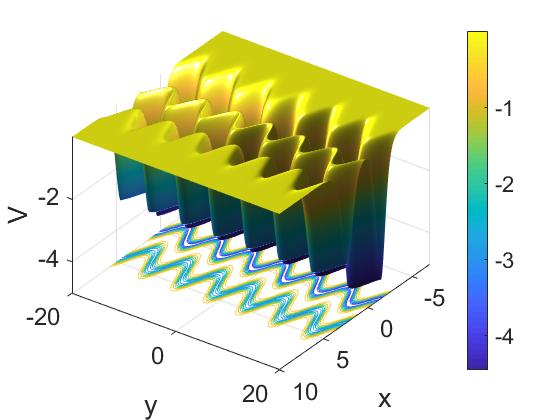}}
\subfigure[$\Phi(y)=\sech(y)$]{\includegraphics[height=2.5cm,width=3.5cm]{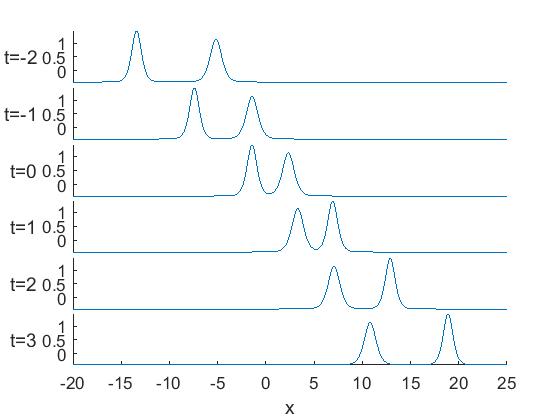}}
\subfigure[$\Phi(y)=\sech(y)$]{\includegraphics[height=2.5cm,width=3.5cm]{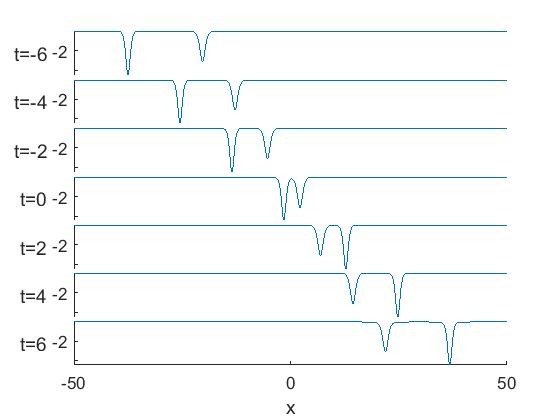}}
\subfigure[$\Phi(y)=\sin(y)$]{\includegraphics[height=2.5cm,width=3.5cm]{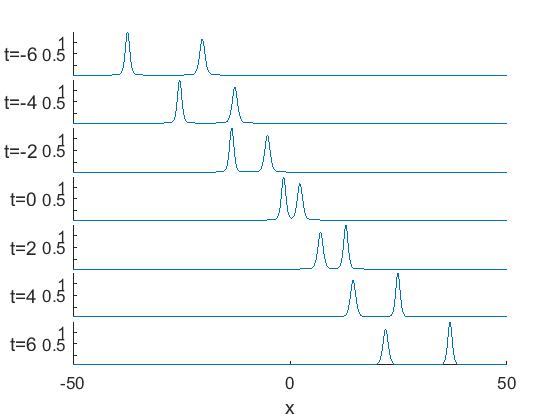}}
\subfigure[$\Phi(y)=\sin(y)$]{\includegraphics[height=2.5cm,width=3.5cm]{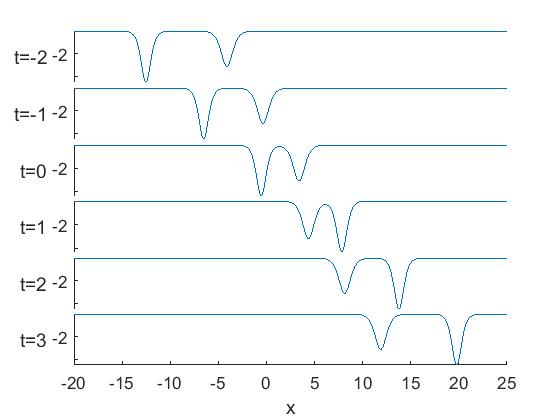}}
\caption{Deformed two-soliton solutions $U_{[2]}$ and $V_{[2]}$ of equation\eqref{N1}
in the $(x,y)$-plane displayed at $t=0$. Panels (a),(b),(c),(d) are the
two-dimensional plots of (e),(f),(g),(h) respectively }\label{fig2}
\end{figure}

In order to obtain deformed two-solitons of 2D KdV equation\eqref{N1}, taking
\begin{equation}\label{u2}
\begin{aligned}
&U_{[2]}=-2\partial_{xy}\ln(1+e^{\eta_1}+e^{\eta_2}+e^{\eta_1+\eta_2+A_{12}})+\phi(y),\\
&V_{[2]}=-2\partial_{xx} \ln(1+e^{\eta_1}+e^{\eta_2}+e^{\eta_1+\eta_2+A_{12}}),
\end{aligned}
\end{equation}
where $\eta_1=p_{1}x+\Phi(y)q_{1}y+\Omega_{1}t+\eta^{0}_{1}$ and $\eta_2=p_{2}x+\Phi(y)q_{2}y+\Omega_{2}t+\eta^{0}_{2}$. Substituting \eqref{u2} into \eqref{N1}, the following relations are obtained
\begin{equation}
\begin{aligned}
\Omega_{1}&=-\frac{p_1^2(p_1q_1-3)}{q_1}, \quad \Omega_{2}=-\frac{p_2^2(p_2q_2-3)}{q_2}, \\
\exp(A_{12})&=\frac{(p_1q_2-p_2q_1)^2+p_1p_2q_1q_2(p_1-p_2)(q_1-q_2)}
{(p_1q_2-p_2q_1)^2+p_1p_2q_1q_2(p_1+p_2)(q_1+q_2)}.
\end{aligned}
\end{equation}
Further, taking $p_1=3, q_1=3, p_2=\frac{5}{2}, q_2=3, \eta^0_1=-\eta^0_2=\frac{\pi}{2}$,
the analytical expressions $U_{[2]}$ and $V_{[2]}$ of deformed two-soliton solutions are as follows
\begin{equation}\label{2s}
\begin{aligned}
U_{[2]}&=\frac{\begin{Bmatrix}&\cosh(2\kappa_{11})-\sinh(2\kappa_{11})+982081\left(\sum\limits_{j=2}^{4}
[\cosh(2\kappa_{1j})+\sinh(2\kappa_{1j})]\right )\\&+\left(\sum\limits_{m=1}^{5}
\Gamma_m[\cosh(\kappa_{m})-(-1)^{m}\sinh(\kappa_{m})]\right )\end{Bmatrix}}
{\frac{1}{\phi(y)}\left(\cosh(\kappa_{11})-\sinh(\kappa_{11})+991\sum\limits_{j=2}^{4}[\cosh(\kappa_{1j})
+\sinh(\kappa_{1j})]\right )^2},\\
V_{[2]}&=\frac{\begin{Bmatrix}&\left[\cosh(3\Phi(y))+\sinh(3\Phi(y))\right ](\sum\limits_{m=1}^{5}
\gamma_m[\cosh(\kappa_{m}-3(-1)^{m+1}\Phi(y))]\\&-\sum\limits_{m=1}^{5}
(-1)^{m}[\sinh(\kappa_{m}-3(-1)^{m+1}\Phi(y))])\end{Bmatrix}}
{-1982\left(\frac{1}{991}[\cosh(\kappa_{11})-\sinh(\kappa_{11})]+\sum\limits_{j=2}^{4}[\cosh(\kappa_{1j})
+\sinh(\kappa_{1j}) ]\right )^2},
\end{aligned}
\end{equation}
where\begin{equation}
\begin{aligned}
\kappa_{11}&=-\frac{11}{2}x-6\Phi(y)+\frac{75}{8}t-\frac{\pi}{2},\quad
\kappa_{12}=\frac{5}{2}x+3\Phi(y)+\frac{69}{8}t,\quad
\kappa_{13}=18t+\frac{\pi}{2},\quad\kappa_{14}=3x+3\Phi(y),\\
 \kappa_{1}&=\kappa_{12}+\kappa_{14},\quad
\kappa_{2}=\kappa_{11}-\kappa_{12}, \quad\kappa_{3}=\kappa_{12}+\kappa_{13},\quad
\kappa_{4}=\kappa_{11}-\kappa_{14},\quad
\kappa_{5}=\kappa_{13}+\kappa_{14},\quad
\Gamma_1=1900738, \\
\Gamma_2&=-15856, \quad\Gamma_3=-12767053,\quad
\Gamma_4=-12883,\quad\Gamma_5=15713296, \quad\gamma_1=1112,\quad
\gamma_2=36, \\
\gamma_3&=24775,\quad\gamma_4=25, \quad\gamma_5=35676.
\end{aligned}
\end{equation}
The dynamic behaviors of deformed two-soliton
solutions are more complex and interesting by choosing the appropriate
parameter $\phi(y)$ see Fig.\ref{fig2}. Furthermore, as can be seen the two-dimensional plots
of Fig.\ref{fig2}, the interaction of deformed two-soliton solutions is an elastic collision.

Similarly, $N$-soliton solutions $U_{[N]}$ and $V_{[N]}$ are given in equation\eqref{s1}
of the 2D KdV equation\eqref{N1}, in which $f$ can be written as follows:
\begin{equation}\label{N7}
\begin{aligned}
f=f_{[N]}=\sum_{\mu=0,1}\exp(\sum_{j<k}^{(N)}\mu_{j}\mu_{k}A_{jk}
+\sum_{j=1}^{N}\mu_{j}\eta_{j}),\; \;
\end{aligned}
\end{equation}
here
\begin{equation}\label{N8}
\begin{aligned}
\eta_{j}=p_{j}x+\Phi(y)q_{j}y+\Omega_{j}t+\eta^{0}_{j},
\quad \Omega_{j}=-\frac{p_j^2(p_jq_j-3)}{q_j},\\ \exp(A_{jk})=\frac{(p_jq_k-p_kq_j)^2+p_jp_kq_jq_k(p_j-p_k)(q_j-q_k)}
{(p_jq_k-p_kq_j)^2+p_jp_kq_jq_k(p_j+p_k)(q_j+q_k)},
\end{aligned}
\end{equation}
where $p_{j},q_{j}$ are arbitrary real parameters, $\eta^{0}_{j}$ is
a complex constant, $\phi(y)$ is an arbitrary function of $y$, and the subscript $j$ denotes
an integer. For higher order soliton solutions, more deformed solitons will be generated
from equation\eqref{N7} with appropriate parameter $\phi(y)$, and its
dynamic behavior will be more complex, we did not discuss it here.

\section{Deformed 2D RW of the 2D KdV equation} \label{3}
In order to obtain the deformed 2D RW solutions of the 2D KdV
equation\eqref{N1}, taking
\begin{equation} \label{r1}
\begin{aligned}
q_{1}=\lambda_{1}p_{1}, \quad q_{2}=\lambda_{2}p_{2},\quad \eta_{1}^{0}=\eta_{2}^{0}=i\pi,
\end{aligned}
\end{equation}
in equation\eqref{u2} and take a suitable limit as $p_{1}, p_2\rightarrow 0$,
then we further take $\lambda_{1}=a+bi, \lambda_{2}=a-bi$ in \eqref{r1}. The solutions $U^{[1]}$ and $V^{[1]}$ are given as follows
\begin{equation}\label{rw-1}
\begin{aligned}
U^{[1]}(x,y,t)&=\phi(y)+4b^2\phi(y)\frac{ab^2(a^2+b^2)A_{21}^2-\frac{b^4}{a}A_{22}^2-A_{23}}{(A_{11}^2+A_{12}^2+A_{13})^2},\\
V^{[1]}(x,y,t)&=4b^2\frac{b^2(a^2-b^2)(a^2+b^2)^2B_{11}^2-\frac{b^4}{a^2-b^2}B_{12}^2-B_{13}}{(A_{11}^2+A_{12}^2+A_{13})^2},
\end{aligned}
\end{equation}
with
\begin{equation}
\begin{aligned}
A_{11}&=\frac{3b(a^2-b^2)}{a^2+b^2}t+abx+b(a^2+b^2)\Phi(y), \quad A_{12}=b^2x+\frac{6ab^2}{a^2+b^2}t,\\
A_{13}&=a(a^2+b^2)^2, \quad A_{21}=\frac{a^2+b^2}{a}x+(a^2+b^2)\Phi(y)+3t, \\
A_{22}&=6at+(a^2+b^2)x, A_{23}=a^2(a^2+b^2)^3, \quad B_{11}=\frac{a}{a^2-b^2}x+\Phi(y)+\frac{3}{a^2-b^2}t,\\
B_{12}&=6at+(a^2+b^2)x,\quad B_{13}=a(a^2+b^2)^3.
\end{aligned}
\end{equation}
As can be seen from the above expression, for ensure that
the above solutions $U^{[1]}$ and $V^{[1]}$ are smooth, parameter $a>0$ must be held.

\subsection{Fundamental rational solutions}
\begin{figure}[!htbp]
\centering
\subfigure[t=0]{\includegraphics[height=2.5cm,width=5cm]{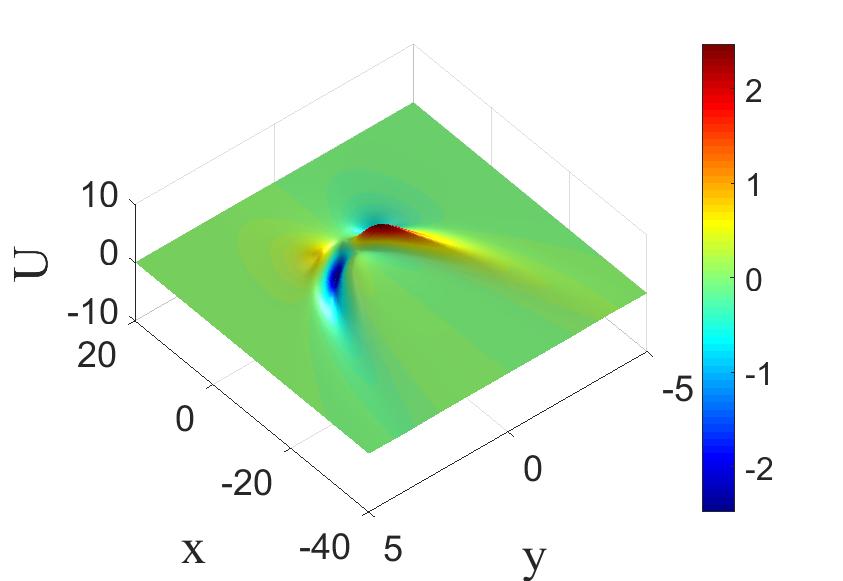}}
\subfigure[t=5]{\includegraphics[height=2.5cm,width=5cm]{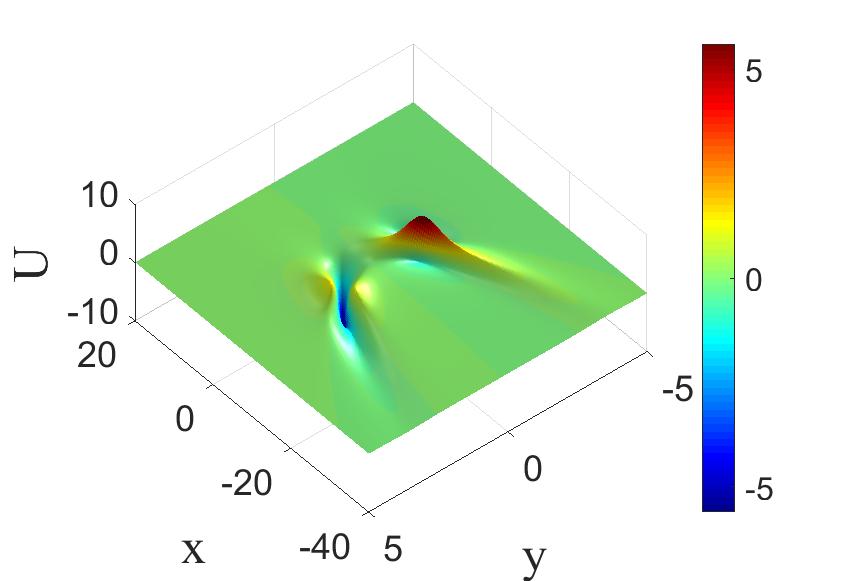}}
\subfigure[t=10]{\includegraphics[height=2.5cm,width=5cm]{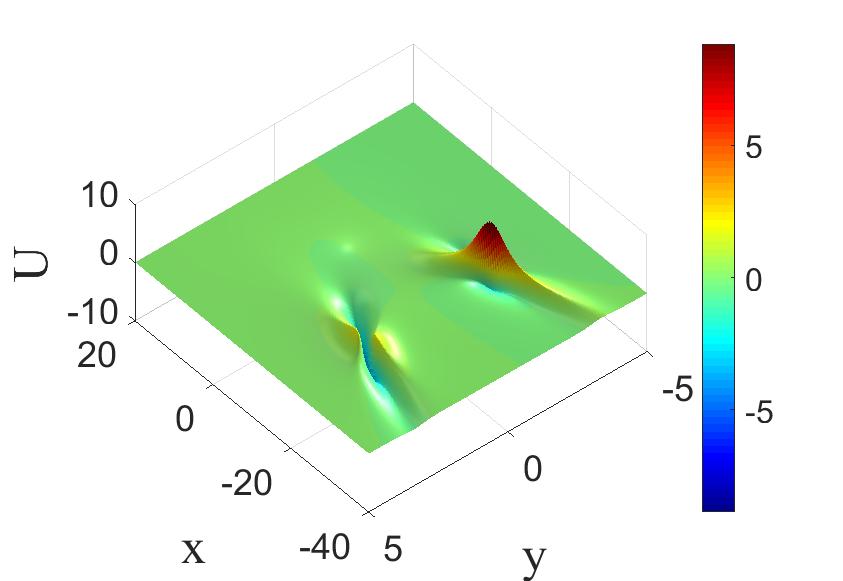}}\\
\subfigure[]{\includegraphics[height=4cm,width=5cm]{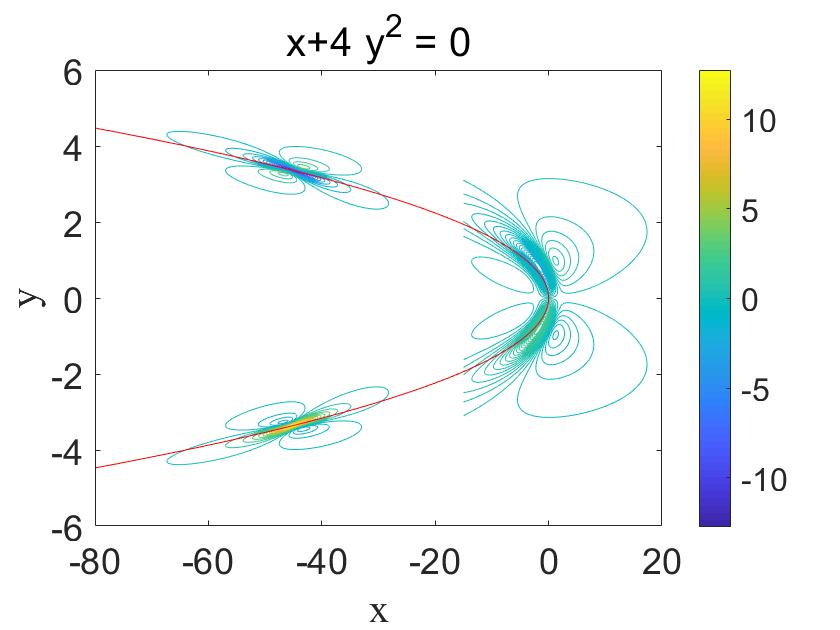}}
\subfigure[t=10]{\includegraphics[height=4cm,width=5cm]{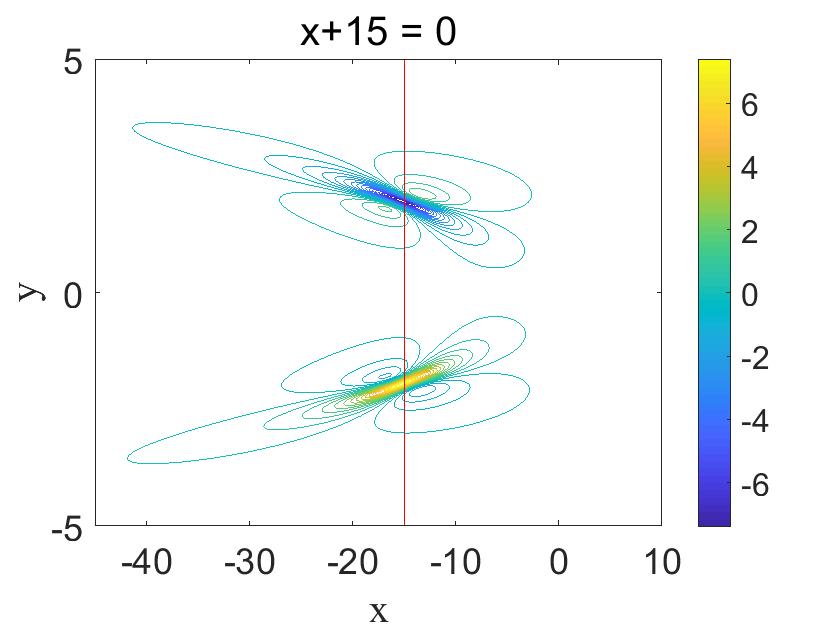}}
\caption{The temporal evolution of fundamental rational solution $U_{[1]}$ removing
the background plane $\phi(y)$ of the equation\eqref{N1} in the $(x,y)$-plane with parameters
$\phi(y)=2y, a=2$ and $b=2$; Panels (d) and (e) are two-dimensional
plots of $U_{[1]}$.}\label{fig3a}
\end{figure}

The fundamental rational solutions $U^{[1]}$ and $V^{[1]}$ of 2D KdV equation\eqref{N1}
are derived if $\phi(y)$ is a polynomial function. Without loss of generality, take
$\phi(y)=2y$. The trajectories of $U^{[1]}$ and $V^{[1]}$ are as follows
\begin{equation}
\begin{aligned}
l_{1}&=ax+(a^2+b^2)y^2+\frac{3(a^2-b^2)}{a^2+b^2}t, \\
l_{2}&=x+\frac{6a}{a^2+b^2}t.
\end{aligned}
\end{equation}
$\lim\limits_{(x,y)\rightarrow\infty}U^{[1]}=2y$ and $\lim\limits_{(x,y)\rightarrow\infty}V^{[1]}=0$,
which show that the background planes of $U^{[1]}$ and $V^{[1]}$ are $2y$ and 0, respectively.
Further, taking $a=2$ and $b=2$, the exact expression of fundamental rational solutions $U^{[1]}$
and $V^{[1]}$ can be obtained. In order to describe the evolution process of
fundamental rational solutions $U^{[1]}$ more clearly,
we removed the background plane $2y$ when plotting. As shown in Fig.\ref{fig3a}. the rational solution $U^{[1]}$ removing
the background plane $2y$ appears from a constant plane when $t\ll 0$.
With time evolution an arc line $x=-4y^2$ appears on the constant background around at $t=0$ [see the Fig.\ref{fig3a}(a)],
and finally the arc line fission into a bright lump and a dark lump [see the Fig.\ref{fig3a}(c)].
The trajectories of fission are $x+4y^2=0$ and $2x+3t=0$ [see the Fig.\ref{fig3a}(d)(e)].

However, the dynamic behavior of the rational solution $V^{[1]}$ is different from $U^{[1]}$. For given
$\phi(y)=2y$, the solution $V^{[1]}$ has the following nine critical points in $(x,y)$-plane
\begin{equation}
\begin{aligned}
&\Lambda_1=(x_1,y_1)=(\frac{3at}{a^2+b^2},y=0);\\
&\Lambda_{2,3}=(x_{2,3},y_{2,3})=(\frac{-3abt\pm\sqrt{3a(a^2+b^2)^3+27b^4t^2}}{b(a^2+b^2)},y=0);\\
&\Lambda_{4,5}=(x_{4,5},y_{4,5})=(-\frac{6at}{a^2+b^2},y=\pm\sqrt{\frac{3t}{a^2+b^2}});\\
&\Lambda_{6,7,8,9}=(x_{6,7,8,9},y_{6,7,8,9})=(\frac{-6abt\pm\sqrt{3a(a^2+b^2)^3}}{b(a^2+b^2)},
y=\pm\sqrt{\frac{3t}{a^2+b^2}}),
\end{aligned}
\end{equation}
letting
\begin{equation}
\begin{aligned}
&\triangle(x,y)=\frac{\partial^2V^{[1]}}{\partial x^2}; \quad
\mathbf{H}(x,y)=\frac{\partial^2V^{[1]}}{\partial x^2}\frac{\partial^2V^{[1]}}{\partial y^2}
-(\frac{\partial^2V^{[1]}}{\partial x\partial y})^2,
\end{aligned}
\end{equation}
yield
\begin{equation}
\begin{aligned}
&\triangle(x,y)\Bigg{|}_{\Lambda_1}=\frac{24b^4(a^2+b^2)^4}{\left[a(a^2+b^2)^3+9b^4t^2\right]^2};\qquad
\mathbf{H}(x,y)\Bigg{|}_{\Lambda_1}=-\frac{1152b^{10}(a^2+b^2)^7t}{\left[a(a^2+b^2)^3+9b^4t^2\right]^4},\\
&\triangle(x,y)\Bigg{|}_{\Lambda_{2,3}}=-\frac{3b^4(a^2+b^2)^4}{4[a(a^2+b^2)^3+9b^4t^2]^2};\quad
\mathbf{H}(x,y)\Bigg{|}_{\Lambda_{2,3}}=-\frac{9b^{10}(a^2+b^2)^7t}{2[a(a^2+b^2)^3+9b^4t^2]^4},\\
&\triangle(x,y)\Bigg{|}_{\Lambda_{4,5}}=\frac{24b^4}{a^2(a^2+b^2)^2};\qquad \qquad \qquad
\mathbf{H}(x,y)\Bigg{|}_{\Lambda_{4,5}}=\frac{2304b^10t}{a^4(a^2+b^2)^5},\\
&\triangle(x,y)\Bigg{|}_{\Lambda_{6,7,8,9}}=-\frac{3b^4}{4a^2(a^2+b^2)^2};\qquad \qquad
\mathbf{H}(x,y)\Bigg{|}_{\Lambda_{6,7,8,9}}=\frac{9b^{10}t}{a^4(a^2+b^2)^5}.
\end{aligned}
\end{equation}
Based on the above analysis, the evolution of rational solution $V^{[1]}$ of 2D KdV equation\eqref{N1}
can be divided into the following three stages.

\begin{itemize}
\item (i) When $t<0$, $\lim\limits_{(x,y)\rightarrow\infty}V^{[1]}=0$, which show that rational solution $V^{[1]}$
appears from a constant background. Now the maximum value $V^{[1]}_{max}(x,y)$
and minimum value $V^{[1]}_{min}(x,y)$ of $V^{[1]}$ are obtained at $\Lambda_{2,3}$
and $\Lambda_{1}$
\begin{equation}
\begin{aligned}
&V^{[1]}_{max}(x,y)=V^{[1]}(x,y)\Bigg{|}_{\Lambda_{2,3}}=\frac{b^2(a^2+b^2)^2}{2a(a^2+b^2)^3+18b^4t^2},\\
&V^{[1]}_{min}(x,y)=V^{[1]}(x,y)\Bigg{|}_{\Lambda_{1}}=-\frac{4b^2(a^2+b^2)^2}{a(a^2+b^2)^3+9b^4t^2}.
\end{aligned}
\end{equation}
Obviously, the global extreme values of rational solution $V^{[1]}$ changes with time,
and $$\lim\limits_{t\rightarrow-\infty}V^{[1]}_{max}(x,y)=\lim\limits_{t\rightarrow-\infty}V^{[1]}_{min}(x,y).$$

\item (ii) When $t=0$, rational solution $V^{[1]}$ has three extreme points
\begin{equation*}
\begin{aligned}
A_1(x,y)=(0,0),\quad A_2(x,y)=(\frac{\sqrt{3a^2+3ab^2}}{b},0),\quad A_3(x,y)=(-\frac{\sqrt{3a^2+3ab^2}}{b},0).
\end{aligned}
\end{equation*}
Maximum and minimum values are as follows
\begin{equation}
\begin{aligned}
&V^{[1]}_{max}(x,y)=V^{[1]}(x,y)\Bigg{|}_{A_2}=V^{[1]}(x,y)\Bigg{|}_{A_3}
=\frac{b^2}{2a(a^2+b^2)},\\
&V^{[1]}_{min}(x,y)=V^{[1]}(x,y)\Bigg{|}_{A_{1}}=-\frac{4b^2}{a(a^2+b^2)}.
\end{aligned}
\end{equation}
From the above analysis, we can seen that its dynamic behavior is similar to
the RW in one-dimensional systems. The amplitude of the RW in one-dimensional systems
is three times that of the background plane. However, the amplitude of RW of
2D KdV equation\eqref{N1} is controlled by parameters $a$ and $b$.

\item (iii) When $t>0$,  rational solution $V^{[1]}$ has six extreme points $\Lambda_{4,5}$ and $\Lambda_{6,7,8,9}$.
And it has four maximums and two minimums
\begin{equation}
\begin{aligned}
&V^{[1]}_{max}(x,y)=V^{[1]}(x,y)\Bigg{|}_{\Lambda_{6,7,8,9}}=\frac{b^2}{2a(a^2+b^2)},\\
&V^{[1]}_{min}(x,y)=V^{[1]}(x,y)\Bigg{|}_{\Lambda_{4,5}}=-\frac{4b^2}{a(a^2+b^2)}.
\end{aligned}
\end{equation}
Obviously, the extreme values of $V^{[1]}$ at $t=0$ is equal to the extreme values at $t>0$.
\end{itemize}

\begin{figure}[!htbp]
\centering
\subfigure[t=-50]{\includegraphics[height=2.5cm,width=3.5cm]{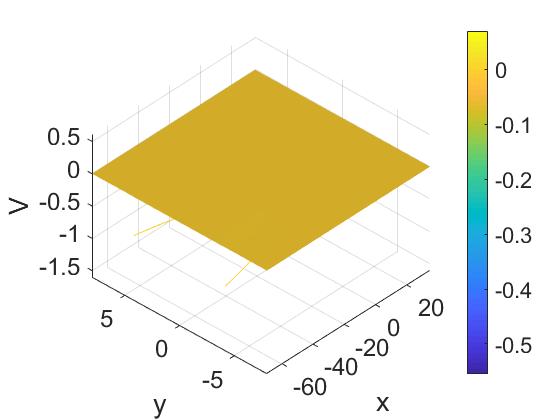}}
\subfigure[t=-9]{\includegraphics[height=2.5cm,width=3.5cm]{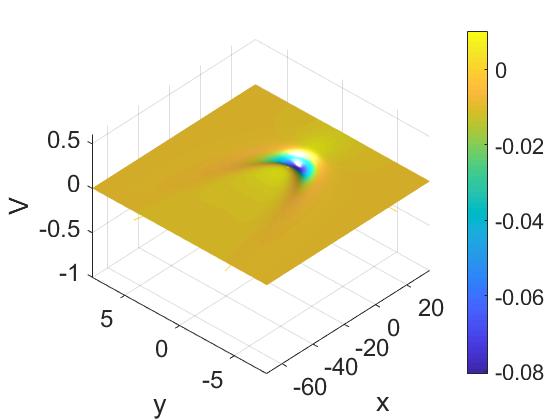}}
\subfigure[t=0]{\includegraphics[height=2.5cm,width=3.5cm]{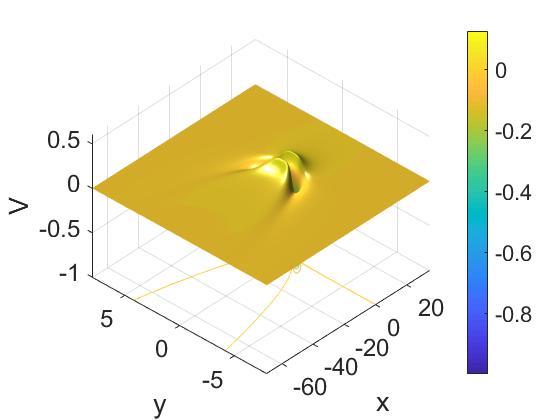}}
\subfigure[t=12]{\includegraphics[height=2.5cm,width=3.5cm]{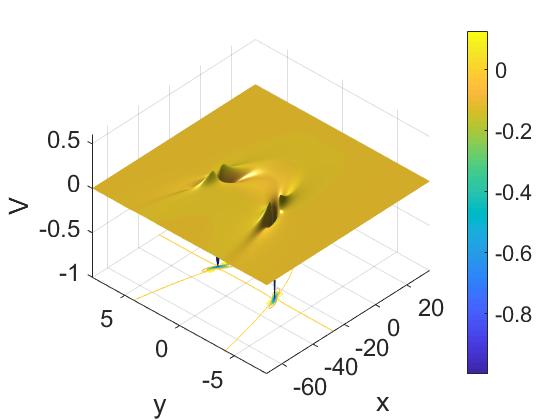}}
\caption{The temporal evolution of fundamental rational solution $V^{[1]}$
of the equation\eqref{N1} in the $(x,y)$-plane with parameters $\phi(y)=2y, a=2$ and $b=2$.}\label{fig3b}
\end{figure}

Through the above analysis, $|V^{[1]}_{max}(x,y)|<|V^{[1]}_{min}(x,y)|$ always satisfied,
which indicated that $V^{[1]}$ is dark. When $t<0$, we know that the rational solution $V^{[1]}$ appears
from a constant plane, and the amplitude above the background plane is $8$ times
of that below the background plane. $V^{[1]}$ reaches the maximum amplitude
$\frac{9b^2}{2a(a^2+b^2)}$ at $t=0$. With the time involution, dark lump
fission into two identical dark lumps along the curve $ax+(a^2+b^2)y^2+\frac{3(a^2-b^2)}{a^2+b^2}t$.
Interestingly, after the dark lump fission into two identical lumps, its amplitude does not change, which means the energy does not change when $t>0$ [see the Fig.\ref{fig3b}].

\subsection{Fundamental deformed 2D RW solutions}
\begin{figure}[!htbp]
\centering
\subfigure[t=-30]{\includegraphics[height=2.5cm,width=3.5cm]{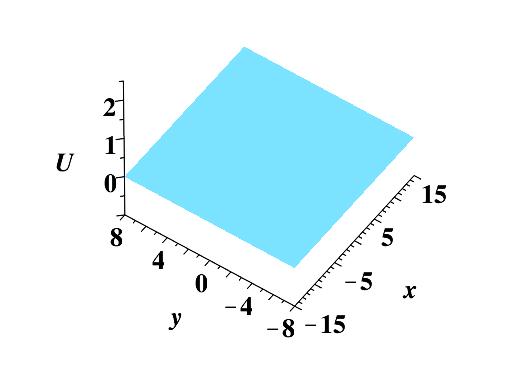}}
\subfigure[t=-2]{\includegraphics[height=2.5cm,width=3.5cm]{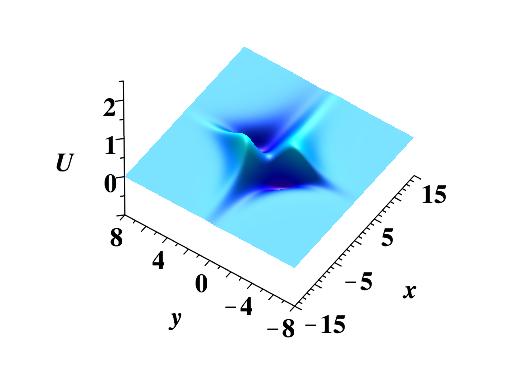}}
\subfigure[t=3]{\includegraphics[height=2.5cm,width=3.5cm]{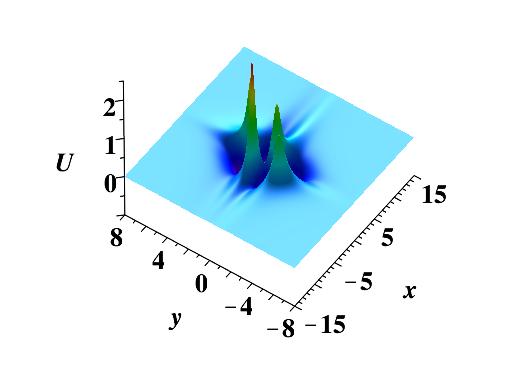}}
\subfigure[t=30]{\includegraphics[height=2.5cm,width=3.5cm]{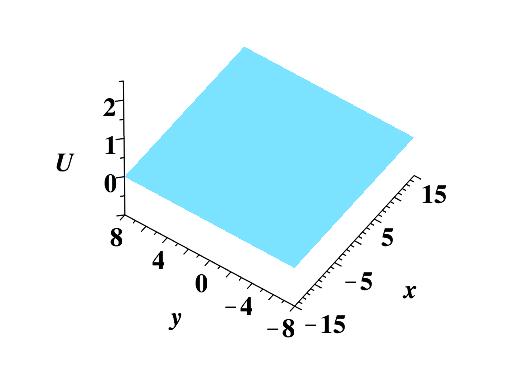}}\\
\subfigure[t=-2]{\includegraphics[height=3cm,width=4.5cm]{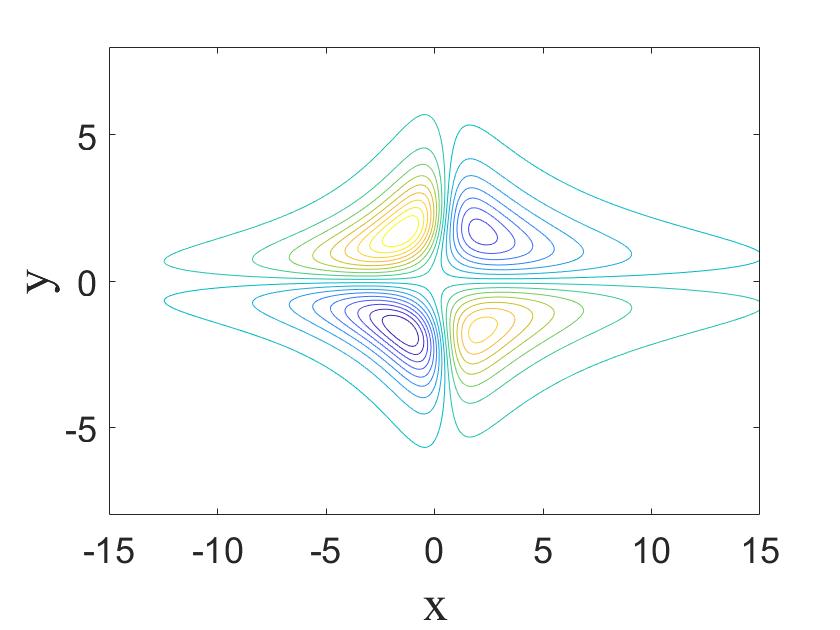}}
\subfigure[t=3]{\includegraphics[height=3cm,width=4.5cm]{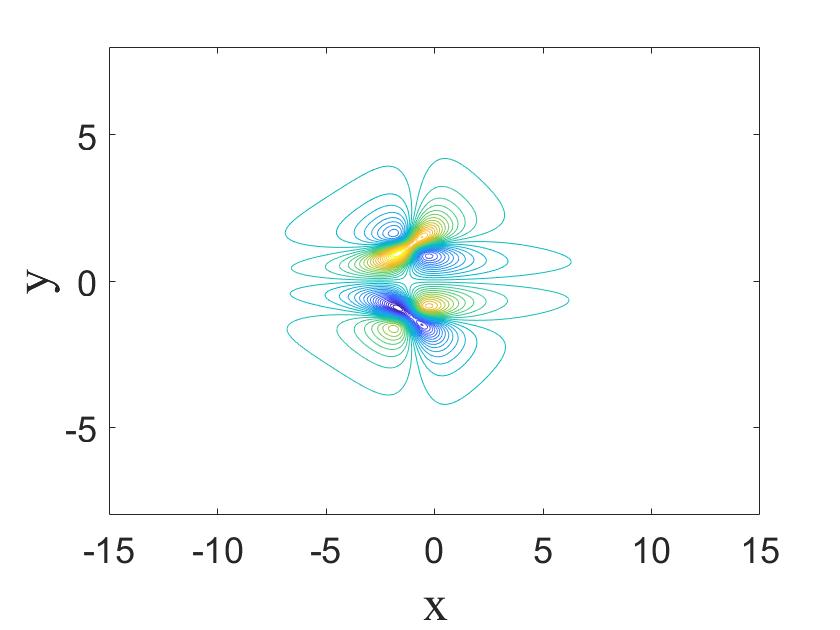}}
\caption{The temporal evolution of fundamental deformed 2D RW solution $U^{[1]}$ removing
the background plane $\phi(y)$ of the equation\eqref{N1} in the $(x,y)$-plane
with parameters $\Phi(y)=\sech(y), a=1$ and $b=4$.}\label{urw1}
\end{figure}

The fundamental deformed 2D RW $U^{[1]}$ and $V^{[1]}$ of 2D KdV equation\eqref{N1}
are obtained when $\phi(y)=\sech(y)$. $\lim\limits_{(x,y)\rightarrow\infty}U^{[1]}=\phi(y)$,
which indicated that the background plane of deformed 2D RW $U^{[1]}$ is $\phi(y)$.
The trajectories of $U^{[1]}$ are
\begin{equation}
\begin{aligned}
l^{'}_{1}&=ax+(a^2-b^2)\sech(y)+3t, \\
l^{'}_{2}&=x+2a\sech(y).
\end{aligned}
\end{equation}
As can be seen in Fig.\ref{urw1}, in order to better observe the evolution
of the deformed 2D RW $U^{[1]}$,
we remove the background plane of $U^{[1]}$. Four panels describe the appearance and annihilation
of 2D RW in $(x, y)$-plane along the curve $x+2\sech(y)$.
This is the first time to obtain such deformed 2D RW in high-dimensional systems.
The dynamic behavior of fundamental deformed 2D RW $V^{[1]}$ is more complicated and interesting.
Through simple calculation and analysis, the evolution of deformed 2D RW $V^{[1]}$ can be divided into
the following four stages

\begin{figure}[!htbp]
\centering
\subfigure[t=-50]{\includegraphics[height=2.5cm,width=3.5cm]{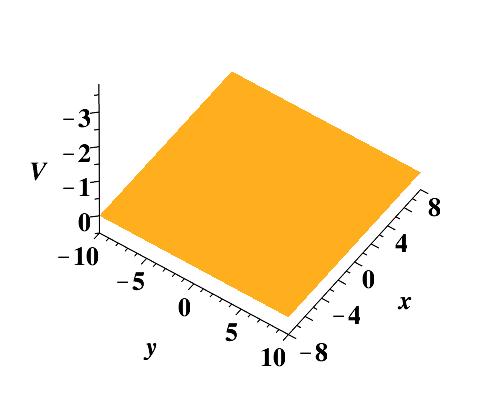}}
\subfigure[t=-6]{\includegraphics[height=2.5cm,width=3.5cm]{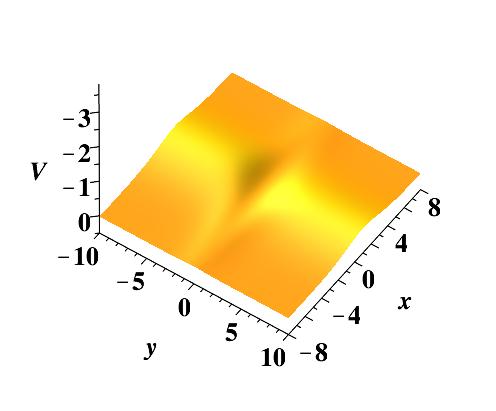}}
\subfigure[t=-2]{\includegraphics[height=2.5cm,width=3.5cm]{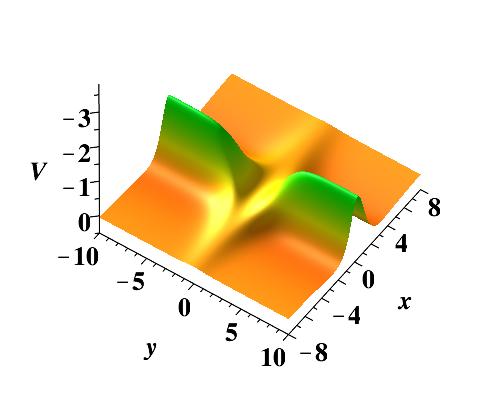}}
\subfigure[t=0]{\includegraphics[height=2.5cm,width=3.5cm]{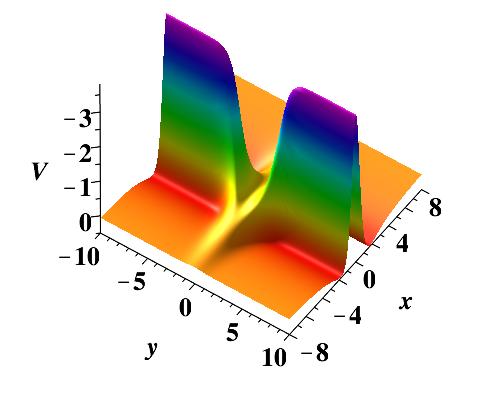}}\\
\subfigure[t=2]{\includegraphics[height=2.5cm,width=4cm]{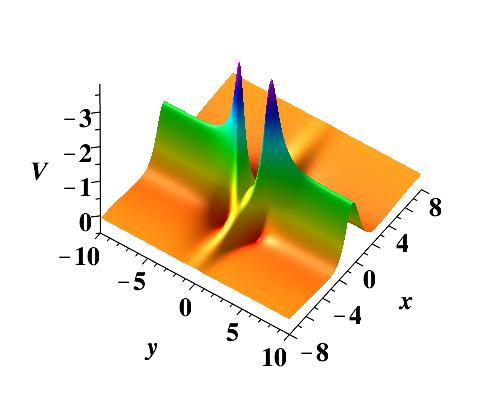}}
\subfigure[t=7]{\includegraphics[height=2.5cm,width=4cm]{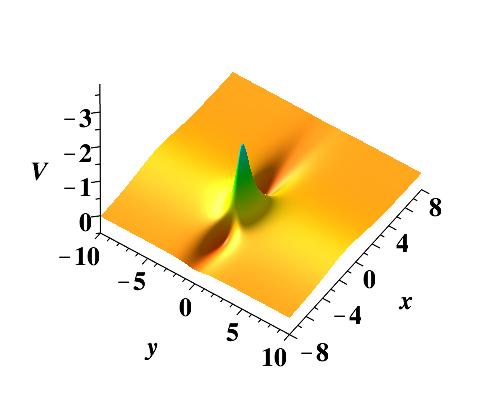}}
\subfigure[t=50]{\includegraphics[height=2.5cm,width=4cm]{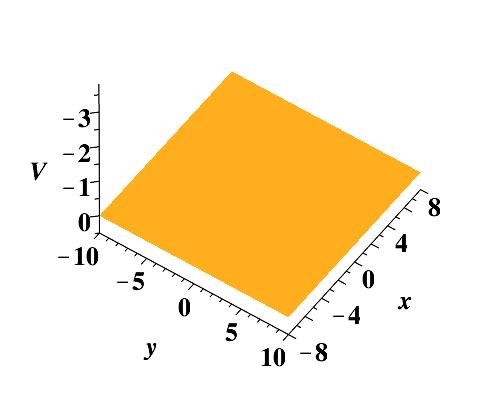}}
\caption{The temporal evolution of fundamental deformed 2D RW solution $V^{[1]}$
of the equation\eqref{N1} in the $(x,y)$-plane with parameters $\Phi(y)=\sech(y), a=1$ and $b=4$.}\label{vrw1}
\end{figure}

\begin{itemize}
\item(i) When $t<0$, a line RW appears from constant background plane, with the time involution,
the amplitude of the line RW gradually increases.

\item(ii) When $t=0$, the amplitude of the line RW reaches its maximum value.
The maximum amplitude $V^{[1]}_{Amp}$ is
\begin{equation}
\begin{aligned}
V^{[1]}_{Amp}=V^{[1]}_{Max}-V^{[1]}_{Min}=\frac{b^2}{2a(a^2+b^2)}-\frac{-4b^2}{a(a^2+b^2)}
=\frac{9b^2}{2a(a^2+b^2)}.
\end{aligned}
\end{equation}

\item(iii) When $0<t\leq\frac{a^2+b^2}{3}$, the line RW disappears gradually. Simultaneously,
a Peregrine-type soliton is generated, but the amplitude of the RW $V^{[1]}$ is fixed
in this process of evolution

\item(iv) When $t>\frac{a^2+b^2}{3}$, two identical maximum values $V^{[1]}_{Max}$ and
one minimum value $V^{[1]}_{Min}$ of the Peregrine-type soliton are as follows
\begin{equation}
\begin{aligned}
V^{[1]}_{Max}=\frac{b^2(a^2+b^2)^2}{2b^4(3t-a^2-b^2)^2+2a(a^2+b^2)^3},\\
V^{[1]}_{Min}=-\frac{4b^2(a^2+b^2)^2}{b^4(3t-a^2-b^2)^2+a(a^2+b^2)^3}.
\end{aligned}
\end{equation}
It can be seen from the above expression that the amplitude of the
Peregrine-type soliton decays rapidly in a very short time. When $t\gg\frac{a^2+b^2}{3}$,
RW $V^{[1]}$ uniformly approaches a constant background see Fig.\ref{vrw1}.
\end{itemize}

\subsection{High-order deformed 2D RW solutions}
\begin{figure}[!htbp]
\centering
\subfigure[t=-20]{\includegraphics[height=2.5cm,width=3.5cm]{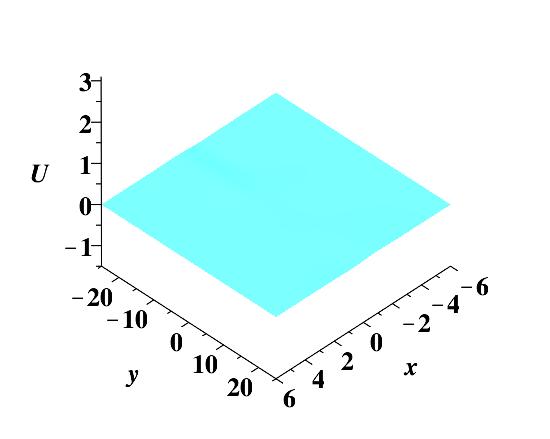}}
\subfigure[t=-2]{\includegraphics[height=2.5cm,width=3.5cm]{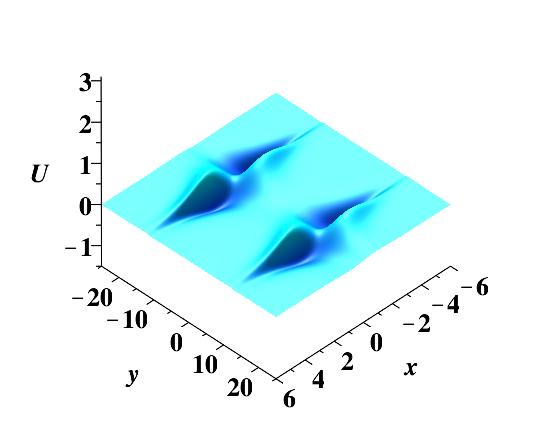}}
\subfigure[t=4]{\includegraphics[height=2.5cm,width=3.5cm]{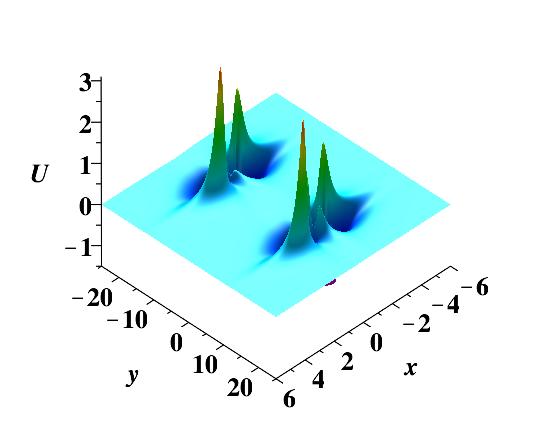}}
\subfigure[t=20]{\includegraphics[height=2.5cm,width=3.5cm]{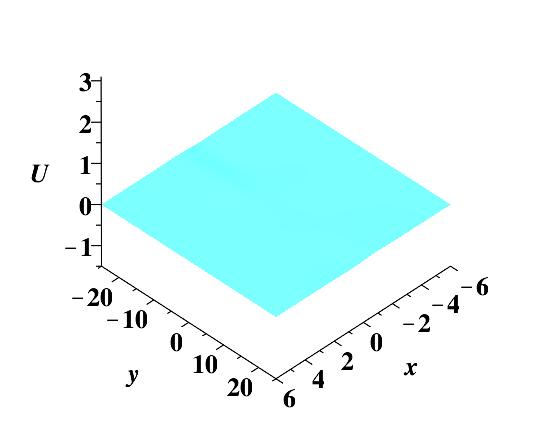}}
\caption{The temporal evolution of second-order deformed 2D RW solution $U^{[2]}$ removing
the background plane $\phi(y)$ of equation\eqref{N1}
in the $(x,y)$-plane.}\label{urw2}
\end{figure}

\begin{figure}[!htbp]
\centering
\subfigure[t=-20]{\includegraphics[height=2.5cm,width=5cm]{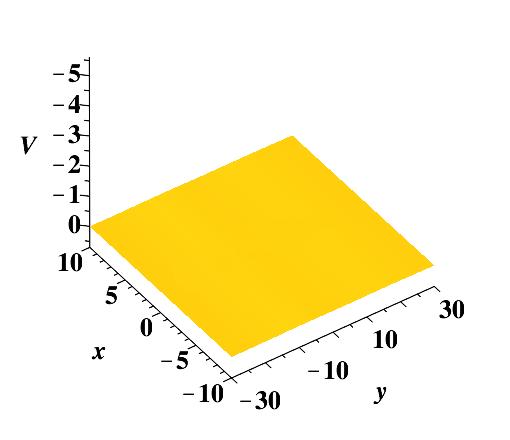}}
\subfigure[t=-2]{\includegraphics[height=2.5cm,width=5cm]{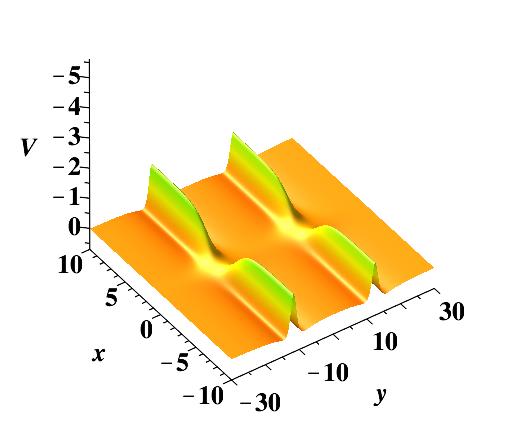}}
\subfigure[t=0]{\includegraphics[height=2.5cm,width=5cm]{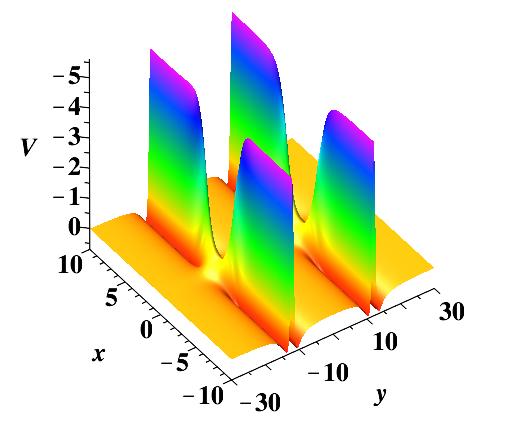}}\\
\subfigure[t=4]{\includegraphics[height=2.5cm,width=5cm]{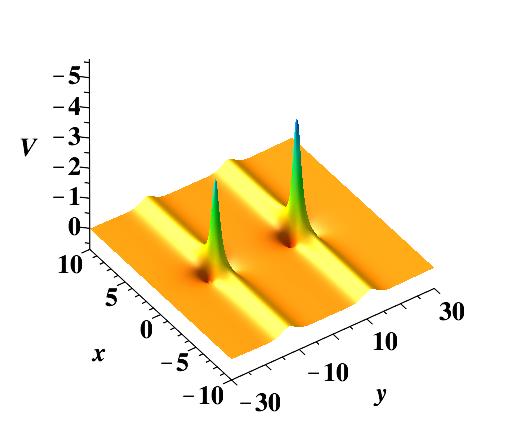}}
\subfigure[t=20]{\includegraphics[height=2.5cm,width=5cm]{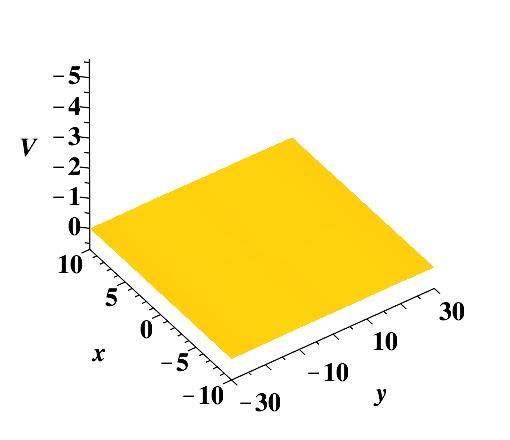}}
\caption{The temporal evolution of second-order deformed 2D RW solution $V^{[2]}$ of equation\eqref{N1} in the $(x,y)$-plane.}\label{vrw2}
\end{figure}

When taking
\begin{equation} \label{r2}
\begin{aligned}
N=4,\quad q_{j}=\lambda_{j}p_{j},\quad \eta_{j}^{0}=i\pi (j=1,2,3,4),
\end{aligned}
\end{equation}
in equation\eqref{N7}. Further taking
\begin{equation}
\begin{aligned}
\lambda_1=\lambda_2^{*}=1+3i,\quad \lambda_3=\lambda_4^{*}=\frac{1}{2}-4i,\quad \Phi(y)=\sech(y),
\end{aligned}
\end{equation}
the second-order deformed 2D RW solutions $U^{[2]}=-2\ln f^{[2]}_{xy}+\phi(y)$
and $V^{[2]}=-2\ln f^{[2]}_{xx}$ are generated,
in which $f^{[2]}$ can be written as,
\begin{equation}\label{2-RW-solution}
\begin{aligned}
f^{[2]}(x,y,t)&=x^4+[6\Phi(y)+\frac{27}{10}t]x^3+[21\Phi(y)^2
+\frac{21}{5}\Phi(y)t+\frac{189}{40}t^2+\frac{149877}{1156}]x^2+
[36\Phi(y)^3+\frac{27}{10}t\Phi(y)^2\\
&+\frac{81}{20}t^2\Phi(y)+\frac{81}{20}t^3+\frac{2924139}{11560}t]x+
\frac{81}{40}t^4-\frac{81}{20}\Phi(y)t^3+[\frac{801}{40}\Phi(y)^2
+\frac{688257}{46240}]t^2+[-\frac{144}{5}\Phi(y)^3\\
&+\frac{1008648}{1445}\Phi(y)]t+40\Phi(y)^4-\frac{122850}{289}\Phi(y)^2
+\frac{1158885}{89}.
\end{aligned}
\end{equation}
As can be seen in Fig.\ref{urw2}, remove the background plane $\phi(y)$,
the second-order deformed 2D RW solution $U^{[2]}$ describes that four Peregrine-type solitons appear
and annihilate from the constant background plane. However, the dynamic behaviors of the second-order
RW solution $V^{[2]}$ are similar to that of the fundamental RW solution $V^{[1]}$.
The RW solution $V^{[2]}$ is uniformly approach to a constant background plane when $t\rightarrow\pm\infty$.
Two line RWs appear from the constant plane under the time evolution,
and their amplitudes increased rapidly. Then the amplitude of the two line RW attenuated rapidly,
at the same time, two Peregrine-type solitons are produced. Finally, these two Peregrine-type solitons
disappeared in a very short time without a trace see Fig.\ref{vrw2}.

For larger N, higher-order solutions are generated, taking the parameters
\begin{equation}
\begin{aligned}
N=2n, \quad q_{j}=\lambda_{j}p_{j}, \quad \eta_{j}^{0}=i\pi\quad (1\leq j \leq N),
\end{aligned}
\end{equation}
in equation\eqref{N7} and take a suitable long wave limit as $p_{j}\rightarrow 0$,
the functions $f$ defined in \eqref{N7} become a polynomial-type
function containing an arbitrary function $\phi(y)$. Therefore, the
general $n$-th rational-type functions $U^{[n]}=-2\ln f_{xy}+\phi(y)$
and $V^{[n]}=-2\ln f_{xx}$ of 2D KdV
equation\eqref{N1} can be derived\cite{ja}, in which $f$ can be written as follows,
\begin{equation}\label{ra-fg1}
\begin{aligned}
f=f^{[n]}=\prod_{k=1}^{N}\theta_{k}+\frac{1}{2}\sum_{k,j}^{(N)}
\alpha_{kj}\prod_{l\neq k,j}^{N}\theta_{l}+\cdots
+\frac{1}{M!2^{M}}\sum_{i,j,...,m,n}^{(N)}\overbrace
{\alpha_{kj}\alpha_{kl}\cdots\alpha_{mn}}^{M}
\prod_{p\neq k,j,...m,n}^{N}\theta_{p}+\cdots,\\
\end{aligned}
\end{equation}
with
\begin{equation}\label{rt}
\begin{aligned}
\theta_{j}=\frac{\lambda_j^2\phi(y)+\lambda_jx+3t}{\lambda_j},
 \quad \alpha_{jk}=-\frac{2\lambda_j\lambda_k
(\lambda_j+\lambda_k)}{(\lambda_j-\lambda_k)^2},
\end{aligned}
\end{equation}
the two positive integers $k$ and $j$. We must emphasize that $\lambda_{j}$ is
a complex constant and $\lambda^{*}_{j}=\lambda_{n+j}=a_n+ib_n$. When $a_n>0$, the above general
$n$-th rational-type functions $U^{[n]}$ and $V^{[n]}$ are smooth.

\noindent\textbf{Remark 1} The above solutions $U^{[n]}$ and $V^{[n]}$ are rational solutions,
if $\phi(y)$ is a nonzero polynomial function. For example, when $\phi(y)=2y$, the rational solution $U^{[n]}$ describes
the fission of $n$-bright lumps and $n$-dark lumps from the background plane $2y$.
The rational solution $V^{[n]}$ describes the fission of $2n$-dark lumps from a constant background plane.

\noindent\textbf{Remark 2} The above solutions $U^{[n]}$ and $V^{[n]}$ are deformed 2D RW solutions,
if $\Phi(y)=\sech(y)$. The solution $U^{[n]}$ describes $2n$ Peregrine-type solitons
appear and annihilate from a kink-soliton plan. The solution $V^{[n]}$ shows that
$n$-line RWs appear and decay rapidly from a constant background plane, and fission into
$n$ Peregrine-type solitons, and finally disappear in the constant background plane without a trace.
\section{Summary}\label{4}
In this letter, new bilinear B$\rm\ddot{a}$cklund transformation
and lax pair of the 2D KdV equation\eqref{N1}
are derived, which are different from the B$\rm\ddot{a}$cklund transform and
lax pair in Ref.\cite{back1,back2}. N-soliton solutions are presented by means of the
improved Hirota's bilinear method. Deformed soliton and deformed breather solutions of elastic collision
are generated by selecting the appropriate free parameter $\phi(y)$
see Fig.\ref{fig1} and Fig.\ref{fig2}. When $\Phi(y)$ is a non-constant polynomial function,
a family of new rational solutions of the 2D KdV equation are generated using the long wave limit.
When $\Phi(y)$ ia a non-zero constant, the rational solutions $U^{[n]}$ and $V^{[n]}$
reduce to the rational solution of 2D KdV equation in Ref.\cite{wxy}.
When $\Phi(y)=\sech(y)$, the deformed 2D RW solution $U$ describes a family of Peregrine-type solitons appear
and annihilate from a kink-soliton plan. In order to better observe the evolution
of the deformed 2D RW solutions, we remove the background plane of $U$ when
plotting see Figs.\ref{urw1} and \ref{urw2}. The deformed 2D RW solution $V$ shows that a series of
line RWs appear and decay rapidly from a constant background plane, and fission into
Peregrine-type solitons, and finally annihilate in the constant background
plane without a trace see Figs.\ref{vrw1} and \ref{vrw2}. This paper successfully constructed
the deformed 2D RW solutions, which are closely related to the
introduced arbitrary function $\Phi(y)$.
These novel phenomena have never been reported before in
nonlinear systems. Our presented work not only provides a new
reference method for seeking new exact solutions of nonlinear partial
differential equations, but also may be helpful to promote
a deeper understanding of nonlinear phenomena.

\section*{Acknowledgments}
This work is supported by the NSF of China under Grants No.11671219, No.11871446,  No.12071304 and No.12071451.


\end{document}